\documentclass[conference]{IEEEtran}
\IEEEoverridecommandlockouts

\usepackage[T1]{fontenc}
\usepackage[utf8]{inputenc}
\usepackage{cite}
\usepackage{microtype}
\usepackage{amsmath,amssymb,amsfonts}
\usepackage{graphicx}
\usepackage{xcolor}
\usepackage{array}
\usepackage{booktabs}
\usepackage{multirow}
\usepackage{url}
\usepackage[draft]{hyperref}
\title{SLT 2026 REAL-TSE Challenge: \\
Real-World Target Speaker Extraction from Conversational Recordings}

\author{
Shuai Wang$^{1}$, Zihan Qian$^{1}$, Ke Zhang$^{2}$, Jiangyu Han$^{3}$, Zikai Liu$^{4}$, Xiaoyang Yu$^{1}$, Haoyu Li$^{1}$\\
Marc Delcroix$^{5}$, Kai Yu$^{6}$, Lei Xie$^{4}$, Ming Li$^{2}$, and Haizhou Li$^{2}$\\
$^{1}$Nanjing University, China,
$^{2}$Chinese University of Hong Kong (Shenzhen), China\\
$^{3}$Brno University of Technology, Czechia,
$^{4}$Northwestern Polytechnical University, China\\
$^{5}$NTT, Inc., Japan,
$^{6}$Shanghai Jiao Tong University, China\\
realtse.challenge@gmail.com
}

\begin{document}
\maketitle

\begin{abstract}
We introduce the \textbf{REAL-TSE Challenge}\footnote{\url{https://real-tse.github.io/challenge/}}, an IEEE SLT 2026 satellite challenge on target speaker extraction~(TSE) from real conversational recordings. Given a multi-speaker mixture and one or more enrollment utterances from a target speaker, participating systems must recover only the target speech. Unlike simulated read-speech benchmarks, REAL-TSE evaluates Mandarin and English recordings that contain natural overlap, reverberation, noise, channel mismatch, and conversational dynamics. The challenge defines two complementary tracks: an \emph{Online} track for low-latency streaming extraction and an \emph{Offline} track for full-context processing. Systems are evaluated with Token Error Rate~(TER), Speaker Similarity~(SpkSim), DNSMOS, and target-speaker activity F1. This overview covers the task, data, baselines, evaluation protocol, submitted systems, condition-wise findings, and lessons for future real-world TSE.
\end{abstract}

\section{Introduction}
\label{sec:intro}

Target Speaker Extraction~(TSE) aims to isolate a specified speaker from a multi-talker recording using auxiliary cues such as enrollment utterances~\cite{wang2019voicefilter,xu2020spex,zmolikova2023neural}. Robust TSE is an enabling technology for teleconferencing, hearing assistance, smart assistants, and speaker-aware meeting transcription.

Despite substantial progress, most widely used TSE evaluations still rely on simulated mixtures such as LibriMix~\cite{cosentino2020librimix} and WSJ0-2Mix~\cite{hershey2016wsj02mix}, where clean read-speech utterances are summed at predefined signal-to-noise ratios. These benchmarks are reproducible and analytically convenient, but they do not fully capture the conditions that make conversational extraction difficult: uncontrolled loudness variation, real reverberation, ambient noise, turn-taking, reactive overlap, disfluencies, and non-verbal vocalizations. Existing challenges have advanced far-field ASR and speech enhancement~\cite{barker2018chime5,watanabe2020chime6,sutherland2025descriptor,yu2022m2met,fu2021aishell4} in realistic acoustic environments, yet none directly benchmarks real-recorded TSE in both low-latency and full-context settings.

The \textbf{REAL-TSE Challenge} addresses this gap with four design choices. First, it evaluates real Mandarin and English conversational recordings, combining data derived from \textsc{REAL-T}~\cite{realt2025} with an independently collected unseen subset. Second, it separates online and offline tracks so that streaming and batch systems can be assessed under appropriate constraints. Third, it uses a multi-dimensional scoring protocol that covers intelligibility, speaker consistency, perceptual quality, and target-speaker activity. Fourth, it releases baselines and validation tools built on \textsc{WeSep}~\cite{wang2024wesep}\footnote{\url{https://github.com/REAL-TSE/wesep-real-tse}}, together with an official toolkit\footnote{\url{https://github.com/REAL-TSE/REAL-TSE-Challenge}} for submission checking and metric computation. The challenge attracted registrations from more than 100 teams spanning over 15 countries and regions. Ultimately, 18 teams submitted 24 valid systems, with 12 submissions in each track. Their results are available on the Leaderboard.\footnote{\url{https://real-tse.github.io/challenge/\#rankings}} Following the challenge, we further refined the dataset and made it publicly available.\footnote{\url{https://huggingface.co/datasets/REAL-TSE/REAL-T}}

\section{Related Datasets and Challenges}
\label{sec:related}

Simulated TSE sets such as WSJ0-2Mix~\cite{hershey2016wsj02mix} and LibriMix~\cite{cosentino2020librimix} provide clean references but omit natural conversational behavior. Meeting and far-field resources such as CHiME~\cite{barker2018chime5,watanabe2020chime6}, AMI~\cite{carletta2005ami}, AISHELL-4~\cite{fu2021aishell4}, and M2MeT~\cite{yu2022m2met} capture realistic acoustics, but primarily target ASR, diarization, or enhancement rather than enrollment-conditioned TSE.

\textsc{REAL-T}~\cite{realt2025} is the closest prior real-conversation TSE resource. REAL-TSE extends it with a newly recorded bilingual unseen set, controlled device pairings, online/offline tracks, four-metric ranking, and post-challenge diagnostics for reproducible comparison under real conversational conditions.

\section{Challenge Settings}
\label{sec:tracks}

Each REAL-TSE trial consists of a real multi-speaker mixture $\mathbf{x}$ and an enrollment utterance $\mathbf{e}$ from the target speaker. The system outputs an estimate $\hat{\mathbf{s}}$ of the target speech while suppressing non-target speakers and non-speech artifacts.

\subsection{Track 1: Online Target Speaker Extraction}
\label{sec:track1}

The online track targets latency-sensitive applications such as hearing devices and real-time communication. Systems must operate frame-wise or chunk-wise without full-utterance buffering, and their end-to-end algorithmic latency must not exceed 100~ms. The latency budget includes future context introduced by STFT/iSTFT, overlap-add, look-ahead frames, smoothing, post-processing, and chunk accumulation, but excludes hardware computation and I/O time. Effective latency is verified with a perturbation-based response-delay test.

\subsection{Track 2: Offline Target Speaker Extraction}
\label{sec:track2}

The offline track targets batch scenarios such as meeting transcription and audio post-production. Systems may use the full utterance, global context, arbitrary model classes, and unrestricted inference cost. However, each mixture--enrollment pair must still be processed independently, without metadata or information from other evaluation samples.

\subsection{Evaluation Metrics}
\label{sec:metrics}

Systems are ranked by dense ranking on four metrics, and the final rank is the average of the four metric ranks. As discussed in Sections~\ref{sec:discussion} and~\ref{sec:system_lessons}, this multi-objective design improves coverage of TSE quality but can also encourage metric gaming.

\paragraph{Intelligibility: Token Error Rate (TER)}
TER measures ASR transcription accuracy, using word error rate for English and character error rate for Mandarin. Lower TER indicates better preservation of the target speaker's linguistic content. The official ASR backbone is Zipformer~\cite{yao2024zipformer}.

\paragraph{Speaker Consistency: Speaker Similarity (SpkSim)}
SpkSim is the cosine similarity between speaker embeddings of $\hat{\mathbf{s}}$ and $\mathbf{e}$, computed with the WeSpeaker ResNet-34 encoder~\cite{wang2023wespeaker}. Higher scores indicate better target-speaker consistency. Tables and figures abbreviate this metric as SIM.

\paragraph{Perceptual Quality: DNSMOS}
DNSMOS is a non-intrusive perceptual-quality estimator reflecting artifacts, distortion, and residual noise. The challenge initially considered the DNSMOS OVRL~\cite{reddy2022dnsmos} sub-score for perceptual-quality ranking. However, post-submission reliability analysis and subjective listening tests showed that OVRL suffered from metric-specific over-optimization in this competitive setting. We therefore \textbf{adopt DNSMOS-P808~\cite{reddy2021dnsmos} as the official perceptual-quality metric}, which appeared not to have been the target of over-optimization, and report OVRL only for reference. Details are provided in Section~\ref{sec:dnsmos_reliability}.

\paragraph{Target-Speaker Activity: F1 Score}
F1 measures whether speech is emitted when, and only when, the target speaker is active. It combines temporal precision and recall over activity regions detected with FireRedVAD~\cite{xu2026fireredasr2s}.

\paragraph{Evaluation References}
EVAL-2 transcripts and target-activity references are human-annotated from the speaker-specific headset channels described in Section~\ref{sec:evalset}. The close-talk channels resolve the target content and timing during overlap, while FireRedVAD is applied only to determine activity in system outputs. This design avoids relying on the far-field mixture to decide which speaker produced a token or was active during overlap.

\section{Datasets}
\label{sec:data}

Most TSE benchmarks are simulated from clean read speech~\cite{allen1979image}, which simplifies evaluation but introduces a mismatch with real reverberation, loudness variation, noise, and conversational turn-taking. REAL-TSE therefore grounds evaluation in real conversational recordings where overlap, speaker behavior, and channel effects arise naturally.

\subsection{Challenge Splits and Statistics}
\label{sec:split_stats}

Table~\ref{tab:split_overview} summarizes the three challenge splits. Each sample is a released mixture--enrollment pair, and multiple samples may share mixture or enrollment files. In total, REAL-TSE contains 6{,}921 trials over 2{,}290 mixtures and 11.2\,h of mixture audio in Mandarin and English.
\begin{table}[!tb]
\centering \caption{Statistics of the challenge dataset. Samples are mixture-enrollment pairs; mixtures and enrollments are unique WAV counts; target speakers are distinct speakers.}
  \label{tab:split_overview}
\scriptsize \setlength{\tabcolsep}{5pt}
  \begin{tabular}{@{}lrrr@{}}
    \toprule
                          & \textbf{DEV} & \textbf{EVAL-1} & \textbf{EVAL-2} \\
    \midrule
    Samples               & 1{,}991 & 1{,}930 & 3{,}000 \\
    Mixtures              & 528     & 576     & 1{,}186 \\
    Enrollments           & 242     & 177     & 1{,}913 \\
    Target speakers       & 64      & 48      & 77      \\
    Mixture audio (h)     & 2.60    & 2.90    & 5.67    \\
    Enrollment audio (h)  & 0.67    & 0.54    & 4.02    \\
    \bottomrule
  \end{tabular}
\vspace{-1mm}
\end{table}

Table~\ref{tab:tse_metadata} reports statistics for mixture duration, enrollment duration, overlap ratio, and target-speaker activity ratio. 

\begin{table}[t]
\centering
\caption{Mean values of the main characteristics of each data split. The numbers in parentheses show the min and max values.}
\label{tab:tse_metadata}
\scriptsize
\setlength{\tabcolsep}{3pt}
\begin{tabular}{@{}lccc@{}}
\toprule
 \textbf{Quantity} & DEV& EVAL-1& EVAL-2 \\
\midrule
 Mix dur. (s)     & 17.26  (6.24, 29.91)& 18.33 (5.82, 29.99)  & 17.25 (5.48, 29.91)\\
  Enroll dur. (s)  & 9.65  (5.00, 50.51)& 11.00 (5.01, 59.82) & 9.10  (3.00, 42.98)\\
 Overlap ratio    & 0.49   (0.18, 0.99) & 0.48  (0.18, 0.98) & 0.53  (0.18, 0.95)\\
 Target ratio     & 0.75   (0.27, 1.00)  & 0.75  (0.21, 1.00)& 0.73  (0.31, 1.00)\\
\bottomrule
\end{tabular}
\vspace{-1mm}
\end{table}

\subsection{Development Set}
\label{sec:devset}

The development set contains 1{,}991 mixture--enrollment pairs from \textsc{REAL-T}~\cite{realt2025}, built from AISHELL-4~\cite{fu2021aishell4}, AliMeeting~\cite{yu2022m2met}, AMI~\cite{carletta2005ami}, DipCo~\cite{vansegbroeck2020dipco}, and CHiME6~\cite{watanabe2020chime6}. Each sample pairs a naturally overlapping mixture with a non-overlapping target-speaker enrollment segment of at least 5 seconds. The development set is intended only for tuning, comparison, and ablation, and must \textbf{not} be used for training or fine-tuning.

\subsection{Evaluation Set}
\label{sec:evalset}

The evaluation set contains \textbf{4{,}930 mixture--enrollment pairs} in two subsets (Table~\ref{tab:split_overview}).

\paragraph{EVAL-1 (Seen, 1{,}930 pairs)}
EVAL-1\footnote{See Section~\ref{sec:eval-corrections} for details on the evaluation corrections.} is drawn from the same source corpora as the
development set but contains no overlapping samples,
measuring in-domain performance.

\paragraph{EVAL-2 (Unseen, 3{,}000 pairs)}
EVAL-2 is newly collected and covers meetings, caf\'{e}s, homes, and in-vehicle conversations. Each conversation was synchronously recorded with high-quality microphones (H1/H2), a phone, and one close-talk headset per speaker. Human annotators used the clear per-speaker headset channels (which contain little cross-talk) to produce transcripts and target-activity labels, including during overlap. Target speakers also provided external single-speaker enrollment utterances. EVAL-2 tests generalization under unseen environments, devices, and channel mismatch.

\begin{table}[!tb]
\centering \caption{EVAL-2 mixture$\times$enrollment channel pairing (sample counts). H1/H2: high-quality microphones; phone: mobile-phone recording; headset: speaker-worn close-talk headset; external: additional single-speaker enrollment recordings.}
  \label{tab:eval2_cross}
\scriptsize \setlength{\tabcolsep}{5pt}
  \begin{tabular}{@{}l rrrrr r@{}}
    \toprule
    \multirow{2}{*}{\textbf{Mix.$\backslash$Enr.}} & \multicolumn{5}{c}{\textbf{Enrollment channel}} & \multirow{2}{*}{\textbf{Total}} \\
    \cmidrule(lr){2-6}
          & H1 & H2 & phone & headset & external & \\
    \midrule
    H1    & \textbf{397} & 127 & 129 & 141 & 218 & 1{,}012 \\
    H2    & 142 & \textbf{415} & 128 & 145 & 179 & 1{,}009 \\
    phone & 122 & 135 & \textbf{388} & 131 & 203 & 979 \\
    \midrule
    Total & 661 & 677 & 645 & 417 & 600 & 3{,}000 \\
    \bottomrule
  \end{tabular}
\vspace{-1mm}
\end{table}

Table~\ref{tab:eval2_cross} summarizes EVAL-2 channel pairings, including matched- and cross-channel cases used later for channel-mismatch analysis. For a fair blind evaluation, scenario labels, speaker identities, and device metadata are removed from the released samples, and each pair is processed independently.

\subsection{Training Data Policy}
\label{sec:training_data}

REAL-TSE has no official training set. Participants may use open-source data and pretrained models, but they must document all data sources and checkpoints. Development/test splits of AliMeeting, AISHELL-4, AMI, DipCo, and CHiME6 are forbidden for pre-training, training, or augmentation, whereas their official training splits are allowed. This open-training policy encourages diverse data and modeling strategies while exposing the practical upper bound of current approaches.

\section{Baseline Systems}
\label{sec:baseline}

We provide four BSRNN-based baselines~\cite{luo2023bsrnn,zhang_multilevel} implemented with \textsc{WeSep}~\cite{wang2024wesep}. All baselines are trained on fully overlapped Libri2Mix-100~\cite{cosentino2020librimix} at 16\,kHz for 150 epochs. They compare two enrollment-conditioning methods under causal and non-causal extractors, as summarized in Table~\ref{tab:baseline}.

\begin{table}[!tb]
\centering \caption{Baseline model variants.}
  \label{tab:baseline}
\scriptsize \setlength{\tabcolsep}{3pt}
  \begin{tabular}{lll}
    \toprule
    \textbf{Model} & \textbf{Conditioning method} & \textbf{Extractor} \\
    \midrule
    \texttt{BSRNN\_EMB}          & ECAPA-TDNN emb. & Non-causal \\
    \texttt{BSRNN\_EMB\_CAUSAL}  & ECAPA-TDNN emb. & Causal \\
    \texttt{BSRNN\_TFMAP}        & TF-Map + Context & Non-causal \\
    \texttt{BSRNN\_TFMAP\_CAUSAL} & TF-Map + Context & Causal \\
    \bottomrule
  \end{tabular}
  \vspace{-2mm}
\end{table}

\texttt{BSRNN\_EMB} uses a fixed speaker embedding from ECAPA-TDNN~\cite{desplanques2020ecapa,wang2023wespeaker}, whereas \texttt{BSRNN\_TFMAP} combines spectral TF-Map features with contextual embeddings~\cite{zhang_multilevel}.

The causal baselines keep the same conditioning modules but impose causality inside the BSRNN extractor. Specifically, they replace global normalization with feature-axis normalization and bidirectional temporal LSTMs with unidirectional LSTMs.

Table~\ref{tab:baseline_results} reports the official baseline results. Trained only on synthetic English Libri2Mix, they are lower bounds: EVAL TER near or above 0.70 exposes the domain gap and motivates real-data adaptation. The gap also shows why these baselines provide a common reference rather than a competitive recipe.

\begin{table}[!tb]
\centering \caption{Baseline results on DEV, EVAL-1, and EVAL-2. Lower Token Error Rate (TER) is better; higher SIM, DNSMOS P808/OVRL, and F1 are better. Gray numbers denote the previously reported DNSMOS OVRL scores.}
  \label{tab:baseline_results}
\scriptsize \setlength{\tabcolsep}{4pt} \resizebox{\columnwidth}{!}{%
  \begin{tabular}{llccccc}
    \toprule
    \textbf{Set} & \textbf{Model} & \textbf{TER} & \textbf{SIM} & \multicolumn{2}{c}{\textbf{DNSMOS}} & \textbf{F1} \\
    \cmidrule(lr){5-6}
     & & & & \textbf{P808} & \textbf{OVRL} & \\
    \midrule
    \multirow{4}{*}{DEV}
      & \texttt{BSRNN\_EMB}          & 0.695 & 0.501 & 2.82 & \textcolor{gray}{1.66} & 0.841 \\
      & \texttt{BSRNN\_EMB\_CAUSAL}  & 0.703 & 0.492 & 2.69 & \textcolor{gray}{1.63} & 0.829 \\
      & \texttt{BSRNN\_TFMAP}        & 0.703 & 0.521 & 2.72 & \textcolor{gray}{1.50} & 0.838 \\
      & \texttt{BSRNN\_TFMAP\_CAUSAL} & 0.651 & 0.535 & 2.76 & \textcolor{gray}{1.55} & 0.844 \\
    \midrule
    \multirow{4}{*}{EVAL-1}
      & \texttt{BSRNN\_EMB}          & 0.686 & 0.513 & 2.91 & \textcolor{gray}{1.90} & 0.828 \\
      & \texttt{BSRNN\_EMB\_CAUSAL}  & 0.706 & 0.504 & 2.76 & \textcolor{gray}{1.77} & 0.807 \\
      & \texttt{BSRNN\_TFMAP}        & 0.712 & 0.540 & 2.80 & \textcolor{gray}{1.73} & 0.823 \\
      & \texttt{BSRNN\_TFMAP\_CAUSAL} & 0.667 & 0.558 & 2.84 & \textcolor{gray}{1.77} & 0.836 \\
    \midrule
    \multirow{4}{*}{EVAL-2}
      & \texttt{BSRNN\_EMB}          & 0.773 & 0.357 & 2.85 & \textcolor{gray}{1.80} & 0.830 \\
      & \texttt{BSRNN\_EMB\_CAUSAL}  & 0.751 & 0.370 & 2.72 & \textcolor{gray}{1.67} & 0.831 \\
      & \texttt{BSRNN\_TFMAP}        & 0.766 & 0.382 & 2.73 & \textcolor{gray}{1.52} & 0.833 \\
      & \texttt{BSRNN\_TFMAP\_CAUSAL} & 0.742 & 0.391 & 2.75 & \textcolor{gray}{1.59} & 0.835 \\
    \bottomrule
  \end{tabular}
}
\vspace{-2mm}
\end{table}

\section{Overview of Submitted Systems}
\label{sec:discussion}

Across the two tracks, 18 distinct teams contributed 24 valid systems (12 per track). A valid submission complied with the challenge rules, produced valid evaluation outputs, and was accompanied by a system report. Rather than describing each system individually, we distill the recurring design choices that most consistently affected performance. Three of them stand out: (i) how real conversational data is exploited through simulation and adaptation; (ii) how the extractor is designed and conditioned under (or without) the latency budget; and (iii) how training objectives, post-processing, and test-time behavior interact with the automatic metrics.

\begin{figure}[!tb]
    \centering
    \includegraphics[width=\linewidth]{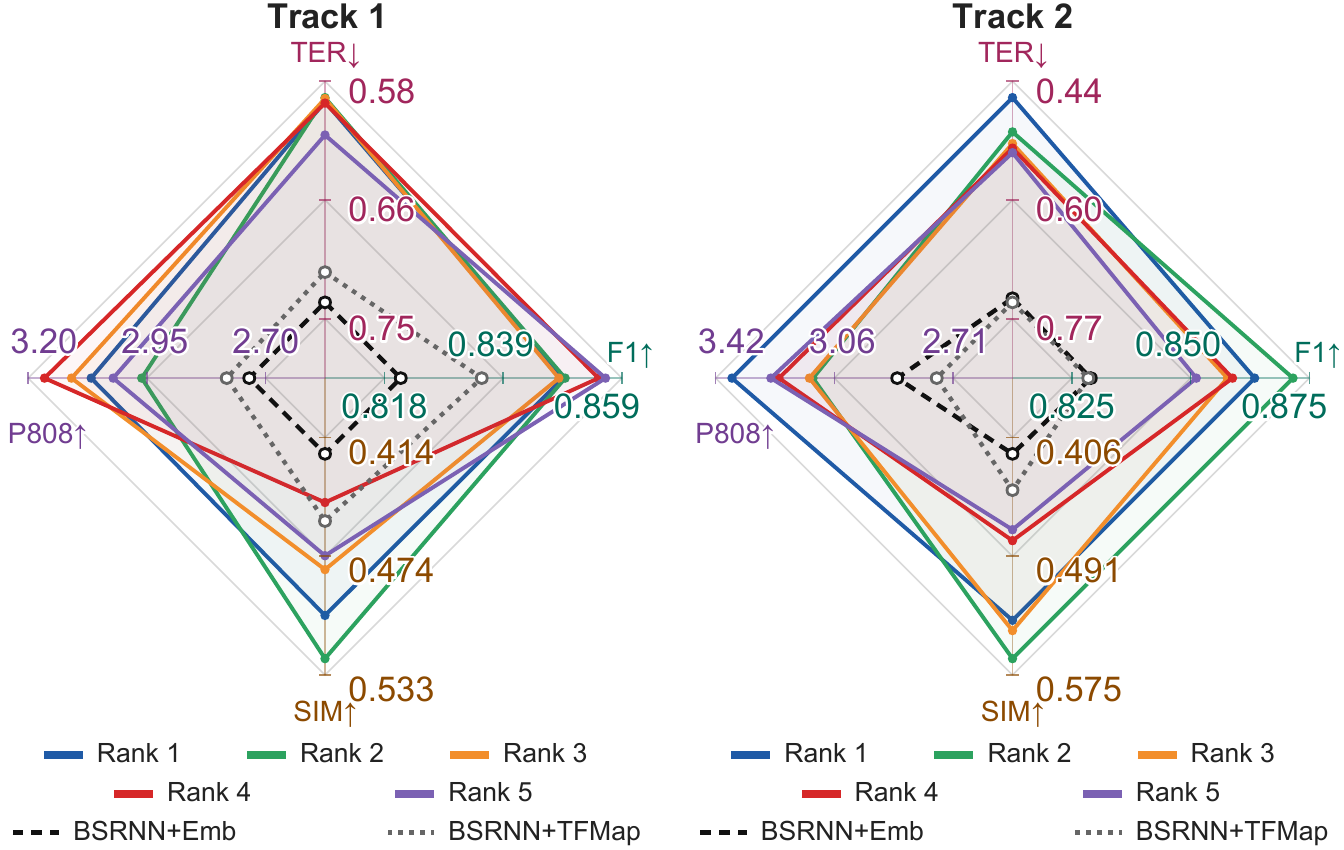}
    \caption{Metric balance of the top-5 valid systems in each track.}
    \label{fig:top5_metric_radar}
    \vspace{-2mm}
\end{figure}

Fig.~\ref{fig:top5_metric_radar} compares the top-five systems in each track with two baseline systems across the four official metrics. All metrics are normalized to a common scale, with TER reversed so that higher values consistently indicate better performance. Top systems substantially outperform both baselines across nearly all metrics. Nevertheless, \textbf{no leading system uniformly dominates across all evaluation dimensions}. Instead, the systems occupy distinct regions of the metric space: some perform better in speech intelligibility or target-speaker activity detection, whereas others more effectively preserve speaker identity or perceptual quality. These complementary strengths highlight the challenging multi-objective nature of real-world TSE, which must simultaneously preserve linguistic content, target activity, speaker identity, and perceptual naturalness.

\subsection{Data Simulation and Real-Data Adaptation}
\label{sec:system_data}

Most teams synthesized mixtures from publicly available speech corpora such as LibriSpeech~\cite{panayotov2015librispeech}/LibriMix~\cite{cosentino2020librimix}, VoxCeleb~\cite{nagrani2017voxceleb}, CN-Celeb~\cite{CN-Celeb}, AISHELL~\cite{ai-shell,fu2021aishell4}, VCTK~\cite{vctk}, and EARS~\cite{ears}; added noise from WHAM!~\cite{wham!}, MUSAN~\cite{musan}, DEMAND~\cite{demand}, or DNS Challenge data~\cite{dnschallenge}; and introduced reverberation with measured~\cite{RIRdataset}/simulated RIRs~\cite{FastRir}, while varying the number of speakers, SNR/SIR, RT60, chunk length, and enrollment quality. Many teams also injected target-absent or noise-only segments, forcing systems not only to extract the target when present but also to suppress non-target speech and avoid false extraction.

The more consequential trend was \textbf{adaptation to real conversational data}. Since real-world mixtures rarely provide clean target references, competitive systems generated pseudo-labels using guided source separation, offline teacher models, model averaging, or self-training. The strongest approaches repeated this process over multiple rounds, gradually progressing from fully overlapped synthetic data to meeting-like mixtures and, ultimately, real far-field recordings. Across the reported results, real-data adaptation consistently outperformed synthetic-only training. Notably, the top three teams either incorporated real recordings into training or relied exclusively on real data.

\subsection{Extractor Design and Speaker Conditioning}
\label{sec:system_modeling}

The latency constraint sharply narrowed the Track~1 design space: nearly all online systems were compact, causal discriminative extractors (mainly BSRNN and TF-GridNet~\cite{wang2023tf} variants) adapted with causal normalization, unidirectional recurrence, and controlled look-ahead. Track~2 explored a much broader space, including SSL-based attractor initialization, diarization-aware front ends, and generative refinement based on diffusion, flow matching, or vocoder reconstruction. Both tracks explored where and how to inject enrollment information through global embeddings, frame-level features, TF-Map conditioning, prefix tokens, or speaker-aware state modulation.

Crucially, \textbf{the results suggest substantial headroom beyond the current submitted systems.} The top-performing entries were all based on BSRNN-style backbones similar to the official baseline, and some used nearly identical extractor architectures, yet achieved substantially better scores. This indicates that realistic data simulation, real-data adaptation, pseudo-label generation and filtering, loss design, metric-aware training, and post-processing can yield large gains even with an existing backbone. At the same time, these results should not be taken to imply that stronger backbones are unnecessary. Rather, they suggest that model architecture and training recipe have not yet been fully co-optimized for real-world TSE. With comparably strong data preparation and adaptation pipelines, more expressive backbones, such as TF-GridNet-style extractors, may therefore still offer considerable room for further improvement.

\subsection{Objectives, Post-Processing, and Metric-Aware Behavior}
\label{sec:system_losses}

Reconstruction losses (SI-SDR/SI-SNR, multi-resolution STFT) were routinely combined with auxiliary terms aligned to the official metrics, including speaker-similarity, DNSMOS-related, ASR/token-level, and VAD/activity losses, marking a clear shift from single-objective separation toward multi-objective optimization. Post-processing followed the same logic, particularly in Track~2: several teams appended enhancement or vocoder-based reconstruction modules, or produced raw and enhanced branches and selected between them using speaker-similarity criteria. Such processing typically improved DNSMOS but could reduce SpkSim or alter target activity, reflecting a trade-off rather than a uniform benefit.

This metric-aware behavior, however, exposed a fundamental weakness of the evaluation itself. Optimizing a neural metric can produce artifacts that raise the score without improving perceived quality. The system reports revealed borderline practices such as branch selection based on automatic scores and repeated use of the official validation scripts during development; in one extreme case, \emph{adversarial waveform perturbations} were applied to inflate SpkSim and DNSMOS without genuinely improving extraction. That submission was ruled invalid and excluded from the official ranking and all valid-system analyses. These cases motivate stricter development rules that \textbf{prohibit use of final non-intrusive ranking metrics} (Section~\ref{sec:system_lessons}).

For Track~1, algorithmic latency mainly came from STFT/iSTFT delay, hop buffering, look-ahead, chunk accumulation, and other forms of non-causal context. Most submitted systems reported latency in the 20--100~ms range. The perturbation-based latency script was important for verifying these claims, since analytical estimates and measured delays did not always agree. In some cases, systems reported substantial look-ahead but showed relatively small measured delay; in others, latency was underestimated because buffering, normalization, or other implementation details were not fully accounted for. Future online TSE benchmarks should require both an analytical latency breakdown and standardized delay measurement.

\section{Condition-wise Analysis of EVAL-2 Results}
\label{sec:result_analysis}

\begin{figure*}[t]
    \centering
    \includegraphics[width=0.85\linewidth]{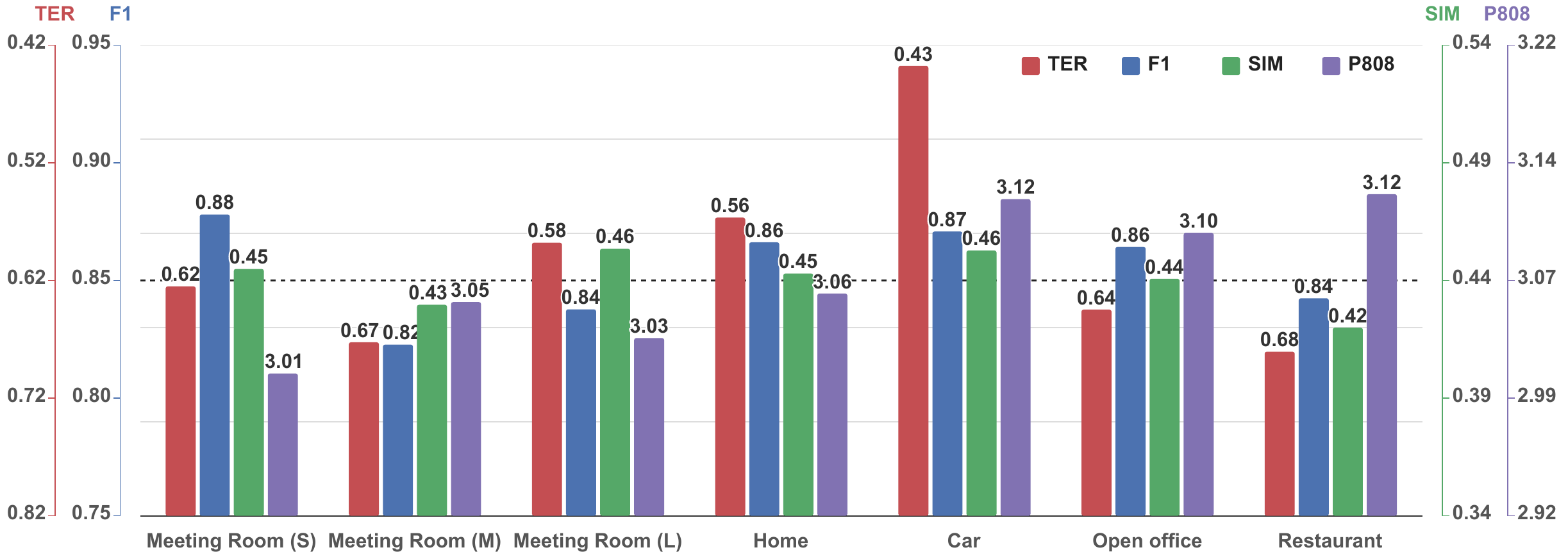}
    \caption{Scenario-wise performance on the EVAL-2 set, computed by averaging all valid submissions separately for CN and EN and then taking a CN/EN macro average over shared scenarios. }
    \label{fig:scenario}
    \vspace{-2mm}
\end{figure*}

Beyond system-report summaries, REAL-TSE supports aggregate analysis of how performance changes under real-world recording conditions. We focus on EVAL-2 because, unlike DEV and EVAL-1, it was newly recorded by the organizers to cover diverse scenarios, devices, and enrollment--mixture channel pairings. We analyze trends by device/channel pairing, target-speaker activity ratio, and scenario.

\subsection{Device and Channel Mismatch Effects}
\label{sec:result_device}
\label{sec:channel_analysis}

\begin{figure}[!tb]
    \centering
    \includegraphics[width=\linewidth]{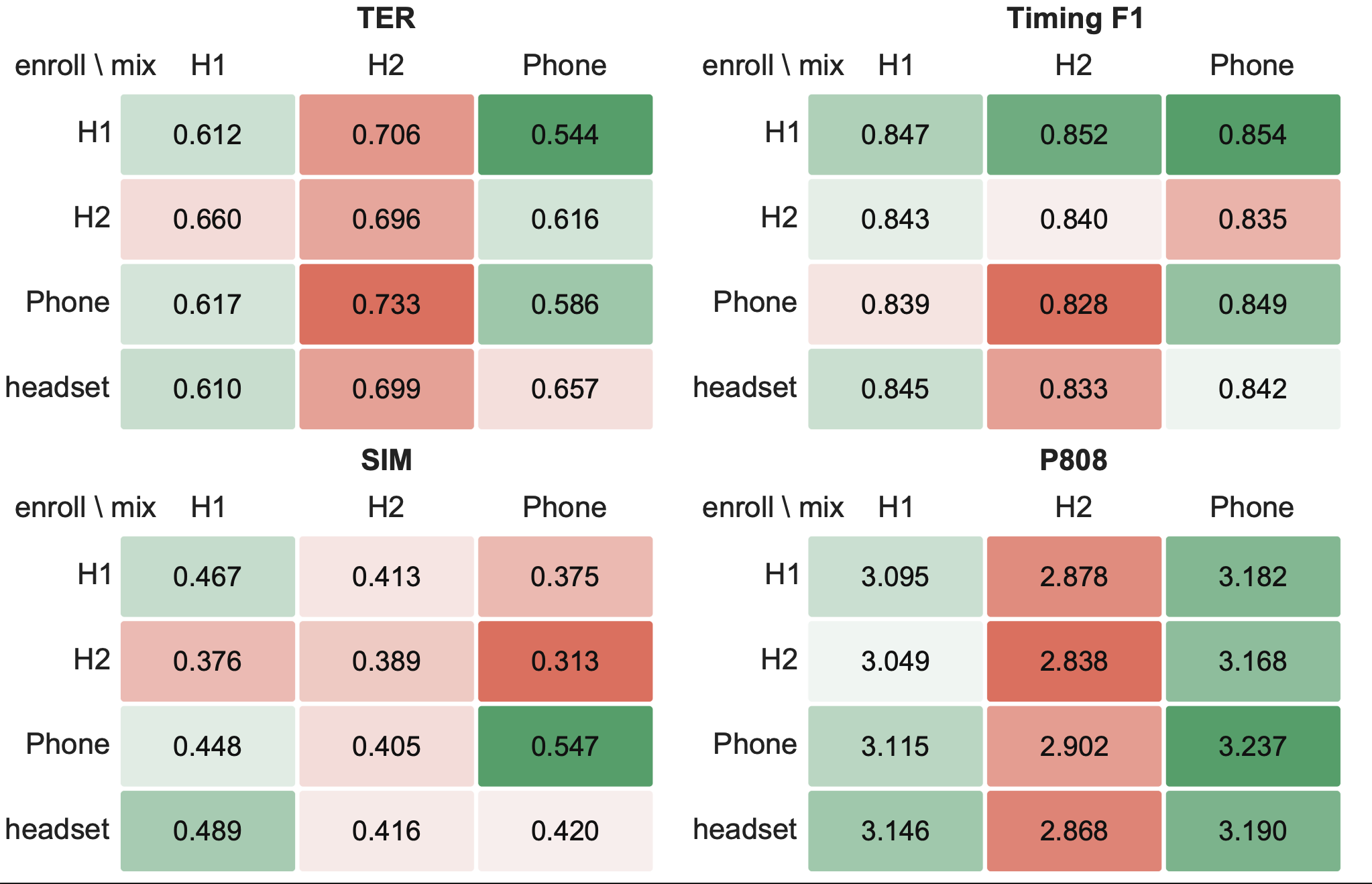}
    \caption{Enrollment-device by mixture-device performance matrix on EVAL-2.}
    \label{fig:channel_matrix}
    \vspace{-2mm}
\end{figure}

We begin by examining channel effects between enrollment and mixture recordings. Figure~\ref{fig:channel_matrix} shows performance measured by the four metrics for different mixture/enrollment microphone pairs, averaged over all valid submitted systems. As shown in Fig.~\ref{fig:channel_matrix}, \textbf{the mixture recording device has a stronger effect than the enrollment device.} Mixtures recorded by the H2 microphone achieved consistently lower performance than H1 and Phone. This matches the recording setup, in which H2 is the farthest microphone and captures more reverberation and environmental interference. Robustness to far-field mixture channels thus remains a key open challenge for real-world TSE. Interestingly, mismatch has different effects across metrics, but systems are generally not over-sensitive except in speaker similarity, which uses enrollment as its reference. External enrollment is omitted from Fig.~\ref{fig:channel_matrix} because it has no corresponding mixture-device row.

\subsection{Target Ratio and Utterance-level Activity}
\label{sec:target_ratio}

\begin{figure}[!htb]
\centering
\includegraphics[width=\linewidth]{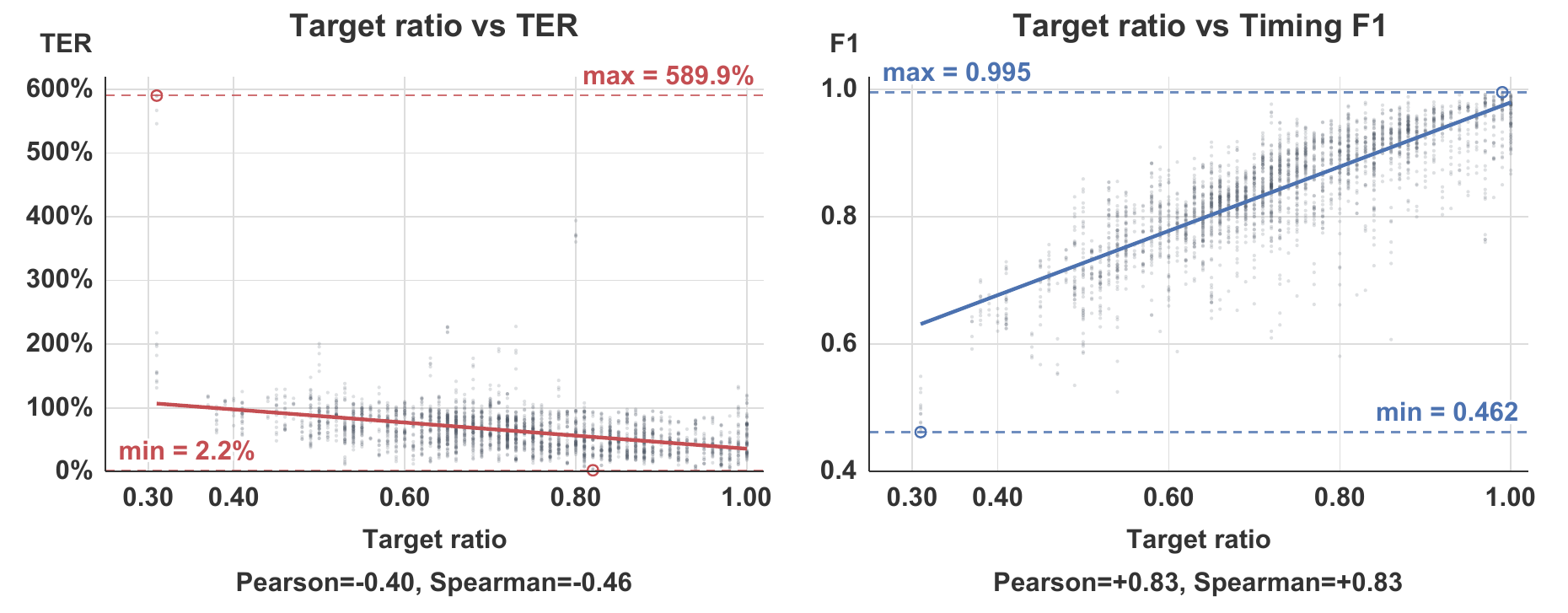}
\caption{Relationship between target ratio and F1/TER on EVAL-2.}
\label{fig:target_ratio}
\vspace{-2mm}
\end{figure}

We further examine the relationship between utterance-level difficulty and the target ratio in Fig.~\ref{fig:target_ratio}, measured as the proportion of target-speaker activity in the mixture.

Overall, target ratio shows the clearest effect on target-activity F1. As the target speaker occupies a larger portion of the mixture, F1 increases, indicating more reliable detection and preservation of target-active regions. TER shows a weaker decreasing trend, suggesting that target ratio alone does not fully explain utterance-level intelligibility errors.

\subsection{Scenario and Conversational Difficulty}
\label{sec:result_scenario}

We next group EVAL-2 samples by recording scenario and compute average performance for each official metric. This tests whether labels for meeting rooms, restaurants, homes, and in-vehicle conversations reliably predict difficulty.

The observed trends do not consistently follow intuitive acoustic expectations. Meeting-room recordings might appear easier, and in-vehicle recordings seem manageable because the car interior is small and the recording geometry is relatively fixed. Conversely, restaurant-like scenarios might be expected to be more difficult because of stronger background noise and less controlled conversational behavior. However, Fig.~\ref{fig:scenario} contradicts this intuitive ordering.

A likely reason is that EVAL-2 conversations were deliberately recorded in a natural and topic-driven manner. Speakers interacted freely rather than following a tightly controlled script, so each scenario contains substantial variability in turn-taking, overlap, speaker distance, loudness, device placement, and local noise events. Scenario-wise statistics are therefore a coarse diagnostic rather than a strict difficulty ranking.

\section{Revisiting DNSMOS Reliability}
\label{sec:dnsmos_reliability}

Near the final stage of the challenge, we observed anomalous behavior in DNSMOS OVRL, the perceptual-quality score originally considered for ranking. Some submissions achieved unusually high OVRL scores that did not align with listening impressions or with other DNSMOS variants.
We therefore conducted a post-submission reliability analysis using the submissions from each track.\footnote{The full analysis and official metric-update notice are available at \url{https://real-tse.github.io/challenge/\#evaluation}. A similar issue was also spotted for speaker similarity, but it appeared to have less impact on evaluation. Due to space constraints, we thus focus our discussion on DNSMOS.}

\begin{figure}[!t]
\centering
\includegraphics[width=\linewidth]{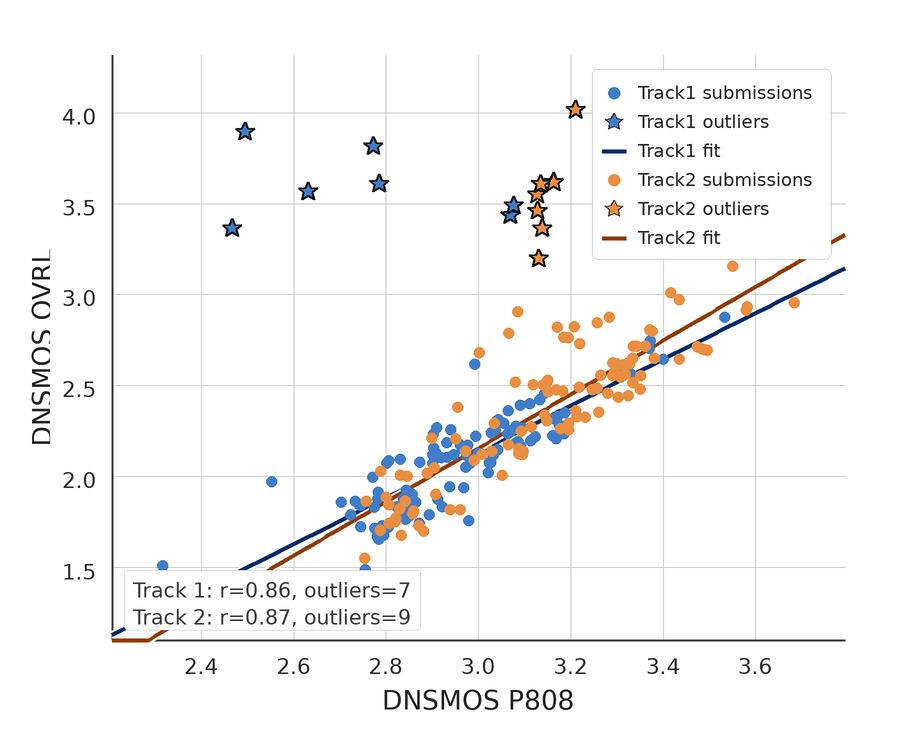}
\vspace{-1mm}
\caption{Relationship between DNSMOS OVRL and DNSMOS P808 on participant submissions. The high-OVRL outliers do not show corresponding gains in P808, indicating metric-specific over-optimization rather than a consistent improvement in perceptual quality.}
\label{fig:dnsmos_ovrl_p808_correlation}
\vspace{-2mm}
\end{figure}

Fig.~\ref{fig:dnsmos_ovrl_p808_correlation} shows the main symptom: several systems obtain high OVRL scores without comparable gains in P808. Since both scores are intended to estimate perceptual quality, such outliers suggest metric-specific over-optimization rather than uniform improvement in extracted-speech quality. To quantify this effect, Table~\ref{tab:dnsmos_correlation} compares DNSMOS variants against human MOS using linear correlation (LCC~\cite{pearson1920notes}) and rank correlation (SRCC~\cite{spearman1961proof}). OVRL is weakly correlated with human MOS, especially in Track~1, whereas P808 shows consistently higher agreement across both tracks. Based on this evidence, we use DNSMOS-P808 as the official perceptual-quality metric and keep OVRL only as a reference score.

\begin{table}[!tb]
\centering
\caption{Correlation between human MOS and DNSMOS variants for top-five final submissions.}
\label{tab:dnsmos_correlation}
\vspace{-1mm}
\scriptsize
\setlength{\tabcolsep}{3pt}
\resizebox{\columnwidth}{!}{%
\begin{tabular}{l c cc cc cc}
\toprule
& & \multicolumn{2}{c}{DNSMOS-OVRL} & \multicolumn{2}{c}{DNSMOS-P808} & \multicolumn{2}{c}{DNSMOS-PRO~\cite{cumlin2024dnsmos}} \\
\cmidrule(lr){3-4}\cmidrule(lr){5-6}\cmidrule(lr){7-8}
Scope & \(n\) & LCC & SRCC & LCC & SRCC & LCC & SRCC \\
\midrule
Track 1 & 500  & +0.003 & +0.040 & +0.453 & +0.380 & +0.156 & +0.115 \\
Track 2 & 600  & +0.182 & +0.186 & +0.466 & +0.463 & +0.229 & +0.234 \\
Overall & 1100 & +0.165 & +0.186 & +0.489 & +0.460 & +0.233 & +0.230 \\
\bottomrule
\end{tabular}%
\vspace{-2mm}
}

\end{table}

These observations highlight a limitation of public reference-free neural metrics: training, post-processing, or selection may over-optimize or exploit them, so higher automatic scores need not indicate better perceived quality.

\section{Overall Lessons}
\label{sec:system_lessons}

The submissions suggest three lessons for future real-world TSE evaluations. First, most teams focused less on proposing fundamentally new extractor architectures and more on making existing backbones work under realistic conditions. Strong systems commonly combined BSRNN- or TF-GridNet-style extractors with careful data simulation, real-data adaptation, pseudo-label filtering, multi-objective training, post-processing, and latency control. This does not imply that architecture is unimportant; rather, backbone design and training recipes need to be co-optimized, and stronger models may still have substantial headroom with equally strong data preparation.

Second, REAL-TSE confirms that TSE quality is inherently multi-dimensional. No system consistently dominates TER, SpkSim, DNSMOS P808, and target-activity F1 at the same time. Different applications may therefore prefer different operating points: meeting transcription may emphasize intelligibility, hearing-assistance and communication scenarios may prioritize latency and perceptual quality, and speaker-aware applications may require stronger identity preservation and activity accuracy. Future leaderboards should report both aggregate rankings and use-case-specific views of performance.

Third, future benchmarks should keep a ranking signal hidden or human-audited, prohibit final-metric optimization or output selection, disclose metric-aware post-processing, and listen to cross-metric outliers. EVAL-2 also shows that fine-grained acoustic and interaction factors explain difficulty better than scenario labels.

\section{Conclusions}
\label{sec:conclusions}

REAL-TSE bridges controlled laboratory benchmarks and real conversational TSE through authentic bilingual recordings, online/offline tracks, and multi-dimensional scoring. The results highlight the joint importance of data construction, speaker conditioning, multi-objective optimization, post-processing, and transparent latency reporting. By releasing the refined dataset, baselines, and evaluation tools, we hope to advance robust, generalizable, and deployable real-world TSE systems.

\section{Post-challenge corrections}
\label{sec:eval-corrections}

During data-release validation and camera-ready preparation, we identified two issues: (i) an audio-channel mapping error in AMI meeting IB4002, previously documented but still present in the source release used,\footnote{\url{https://groups.inf.ed.ac.uk/ami/corpus/dataproblems.shtml}} which caused incorrect speaker labels; and (ii) a text normalization error in our TER evaluation script. We excluded the 70 affected IB4002 samples, leaving 1,930 EVAL-1 samples (4,930 evaluation samples in total), and corrected the normalization. The results reported in this paper incorporate these corrections. Overall rankings remained largely unchanged. Original and corrected rankings are available at \url{https://real-tse.github.io/challenge/#rankings}.

\section*{Acknowledgment}
The REAL-TSE Challenge was sponsored by Huawei. 

This work was supported by the National Natural Science Foundation of China (Grants 62401377 and 62271432), the Yangtze River Delta Science and Technology Innovation Community Joint Research Project (Grant 2024CSJGG1100), and the Shenzhen Science and Technology Program (Shenzhen Key Laboratory, Grant ZDSYS20230626091302006).

\bibliographystyle{IEEEtran}
\bibliography{IEEEbib}

\end{document}